\documentclass[aps, prl, twocolumn, superscriptaddress]{revtex4-2}
\usepackage{graphicx, amsmath, amssymb, color}
\usepackage{xurl, hyperref}
\hypersetup{breaklinks={true}}

\begin{document}
\title{Decentralization can hinder frequency synchronization in power grids through multiple phase transitions} 
\author{Jung-Ho Kim}
\affiliation{Departament d'Enginyeria Inform\`{a}tica i Matem\`{a}tiques, Universitat Rovira i Virgili, 43007 Tarragona, Spain}
\author{Alex Arenas}
\email{alexandre.arenas@urv.cat}
\affiliation{Departament d'Enginyeria Inform\`{a}tica i Matem\`{a}tiques, Universitat Rovira i Virgili, 43007 Tarragona, Spain}
\date{\today}

\begin{abstract}
Decarbonization is rapidly increasing the penetration of inverter-based renewables and other low-capacity generators, intensifying concerns about frequency synchronization in increasingly decentralized power grids.
A common heuristic from Kuramoto onset theory and homogeneous parameter swing-equation models is that distributing generation across many smaller units reduces the effective heterogeneity of nodal injections (natural frequencies) and lowers the coupling required for synchronization.
Here, using a second-order Kuramoto model, we investigate how decentralization affects frequency synchronization when inertia and damping scale with power generation and consumption. We find that decentralization does not always lower the critical frequency synchronization threshold. Instead, increasing decentralization can induce a non-monotonic dependence of the critical coupling strength and lead to a double phase transition in frequency synchronization.
These behaviors remain robust under asymmetric inertia between consumers and generators.
Even when empirical power-generation and power-consumption distributions are considered, a region in which the critical threshold remains nearly constant is observed as decentralization increases.
Our results demonstrate that decentralization can give rise to complex collective dynamics and caution against assuming that decentralization alone ensures improved frequency synchronization.
\end{abstract}
\maketitle

The global effort to mitigate climate crisis is driving a profound transformation of electric power grids, marked by the large-scale integration of renewable energy sources~\cite{IPCC_AR6_WG3_2022,IEA_GridsSecureTransitions_2023}. This transition challenges a foundational pillar of grid stability: frequency synchronization. As synchronous generators are displaced by power-electronic converters, the inherent dynamics that have historically stabilized grid frequency is being fundamentally altered. 
Empirically grounded frequency-fluctuation modeling in island systems illustrates how renewable penetration reshapes frequency dynamics~\cite{Onsaker2023JPhysComplexity,MartinezBarbeito2023IEEETwitterTSE}.
Ensuring the reliability of future grids, therefore, hinges on answering a critical question: how can synchronization be maintained in a fundamentally different power system? Prior work derived explicit stability conditions for synchronous states and identified tunable generator parameters shaping synchronization in realistic grid topologies~\cite{Motter2013NatPhysSpontaneousSynchrony}.

Inverter-based renewable energy sources, exemplified by photovoltaic and wind power plants, are characterized by i) strong fluctuations in power generation due to weather variability~\cite{Widen2015}, ii) low or zero physical inertia~\cite{Ulbig2014}, and iii) a highly decentralized generation structure composed of numerous low-capacity units~\cite{Nadeem2023}. The impacts of the first two characteristics on power-grid dynamics are relatively intuitive. Engineering solutions have been proposed to mitigate their adverse effects on frequency synchronization, such as deploying energy storage systems to balance supply and demand~\cite{Arrigo2020}, and introducing grid-forming inverters to provide virtual inertia~\cite{Fang2018}.
 
However, the collective behavior of frequency synchronization arising from the decentralization of power generation is far from straightforward to understand.
Since decentralization fundamentally alters the network structure of the power grid itself, it represents an inevitable structural feature rather than one that can be mitigated.

In general, power grids can be coarse--grained as networks of second-order Kuramoto oscillators with inertia and damping~\cite{Filatrella2008KuramotoGrid,Nishikawa2015ComparativeModels, 2022WitthautCollective}.
The amount of power generation and consumption at each node enters as a heterogeneous driving term and thus plays the role of a natural frequency in the corresponding second-order Kuramoto formulation~\cite{Dorfler2012,2012RohdenSelf-Organized}.
As a result, the distribution of natural frequencies in power grids is typically bimodal, with positive values for generators and negative values for loads.
When generation is concentrated in a small number of high-capacity units, the separation of these two groups is large.
Conversely, distributing generation across many low-capacity units reduces injection heterogeneity and effectively narrows the bimodal separation.
Since reduced frequency heterogeneity generally promotes synchronization in Kuramoto--type systems, previous studies have reported that increasing decentralization can lower the critical coupling for the onset of synchronization. This conclusion is typically drawn in settings where inertia and damping parameters are held fixed (often taken as homogeneous) while decentralization is varied~\cite{2012RohdenSelf-Organized,Smith2022}.
Recent work also shows that suitably structured heterogeneity (asymmetry) can robustly enhance the stability of synchronous states in real power grids~\cite{Molnar2021NatCommunAsymmetryStability}.

In realistic power systems, the inertia of a conventional generator is proportional to its power capacity~\cite{2020DenholmInertia,Tan2022InertiaReview}, and there exists a positive correlation between inertia and damping~\cite{2019PagnierInertia, 2024LeeReinforcement, 2025ParkOptimal}.
Although inverter-based renewable energy sources possess low or zero physical inertia, the introduction of virtual inertia is a common strategy to compensate for this limitation. Virtual inertia is typically implemented together with (virtual) damping through grid-forming controls~\cite{Kundur1994,ENTSOE2020InertiaRoCoF,Markovic2019}.
In such cases, it is natural to design the virtual inertia to scale proportionally with the power capacity, similarly to conventional generators.

In this letter, we incorporate inertia and damping that scale with the absolute values of power generation and consumption.
Under this more realistic modeling framework, we demonstrate that, in contrast to previous studies, decentralization does not always ensure a lower critical point for frequency synchronization.
Instead, we identify parameter regimes in which increasing decentralization actually hinders frequency synchronization.
In these regimes, a double phase transition in frequency synchronization behavior is observed.
These results remain robust even when the inertia proportionality constant of consumers is set lower than that of generators.
Furthermore, a plateau region in which the critical threshold remains nearly constant as decentralization increases is observed when power-generation and power-consumption distributions derived from empirical data are used.

\vspace{0.1cm}
{\em Second-order Kuramoto networks as a power grid model}\textemdash 
At electromechanical time scales, power balance with inductive coupling yields the network swing equation, a second-order Kuramoto model for grid frequency synchronization~\cite{Dorfler2012}:
\begin{equation}
I_{i}\ddot{{\theta}_{i}} + D_{i}\dot{{\theta}_{i}} = P_{i} + \lambda\sum_{j=1}^{N}A_{ij}\sin({\theta_{j}-\theta_{i}})
\label{Eq:2ndKuramoto}
\end{equation}
Here, $\theta_i$ denotes the phase of node $i$.
$I_i$ and $D_i$ represent the inertia and damping coefficient of node $i$, respectively.
$P_i$ denotes the net power of node $i$, with $P_i > 0$ corresponding to power generation and $P_i < 0$ to power consumption.
$\lambda$ is the coupling strength between nodes, which is assumed to be identical for all links in this model.
$A_{ij}$ is the adjacency matrix of the power-grid network, taking the value 1 if nodes $i$ and $j$ are connected and 0 otherwise, and $N$ is the number of nodes.

While the Kuramoto model is often used to study phase synchronization, phase locking is not required for the operation of power grids.
Instead, stable operation only requires frequency synchronization.
Accordingly, we define the time-averaged frequency order parameter $F$~\cite{2012LozanoRole} as
\begin{equation}
F = \left\langle \frac{1}{N} \sum_{i=1}^{N}\left(\dot{\theta}_i(t)-\langle \dot{\theta}(t) \rangle \right)^2 \right\rangle_t,
\label{Eq:FrequencyOrderParameter}
\end{equation}
where
\begin{equation}
\langle \dot{\theta}(t) \rangle = \frac{1}{N} \sum_{i=1}^{N} \dot{\theta}_i(t)
\end{equation}
denotes the instantaneous average frequency, and $\langle \cdot \rangle_t$ represents a time average taken after the system has reached a steady state.
If $F= 0$, all nodes are frequency synchronized, whereas $F\neq0$ indicates the absence of global frequency synchronization.

Real transmission grids are approximately planar and have sparse connectivity, with mean degree close to three~\cite{2013PaganiThe, SchultzHeitzigKurths2014}. To capture these generic structural features while controlling for topological heterogeneity, we use a two-dimensional honeycomb lattice (mean degree 3) with periodic boundary conditions as a minimal benchmark network. All nodes are assumed to consume a unit amount of power by default, i.e., $P_{i,c} = -1$.
To introduce decentralization, generators are installed at a fraction $p$ of randomly selected nodes.
Each generator produces power $1/p$, such that the power generation balances power consumption,
\begin{equation}
P_{i,g} =
\begin{cases}
1/p, & \text{if node $i$ hosts a generator}, \\
0,   & \text{otherwise}.
\end{cases}
\end{equation}

As a result, the net power at each node is given by
\begin{equation}
P_i = P_{i,c} + P_{i,g},
\end{equation}
so nodes without generators have $P_i = -1$, whereas nodes with generators have $P_i = -1 + 1/p$.
This construction ensures that the network structure remains unchanged as the degree of decentralization $p$ is varied.
This modeling framework represents, for example, a microgrid in which all households consume electricity while only a fraction of them are equipped with local generators, or a high-voltage power grid in which all cities are consumers and only a subset hosts power plants.

Next, the inertia and damping coefficients, which constitute the key dynamical parameters of the model, are assigned to each node as
\begin{equation}
\begin{aligned}
\label{Eq:InertiaDamping}
I_i &= \alpha \left( |P_{i,c}| + |P_{i,g}| \right), \\
D_i &= \beta \left( |P_{i,c}| + |P_{i,g}| \right),
\end{aligned}
\end{equation}
where $\alpha$ and $\beta$ are proportionality constants controlling the strength of inertia and damping, respectively.

Throughout the paper, Eq.~\eqref{Eq:2ndKuramoto} is numerically integrated on a network of $N=1024$ nodes using the fourth-order Runge--Kutta method.
The integration time step is set to $\mathrm{d}t = 0.01$.
After a relaxation time of 1800 time units the frequency order parameter $F$ (Eq.~\eqref{Eq:FrequencyOrderParameter}) is evaluated over an additional time window of 200 time units.
Simulations are initialized with $\theta_i(0)=\dot{\theta}_i(0)=0$. To probe adiabatic response, the final state at a given coupling $\lambda$ is used as the initial condition for the next $\lambda$.

\begin{figure*}[t]
\centering
\includegraphics[width=\textwidth]{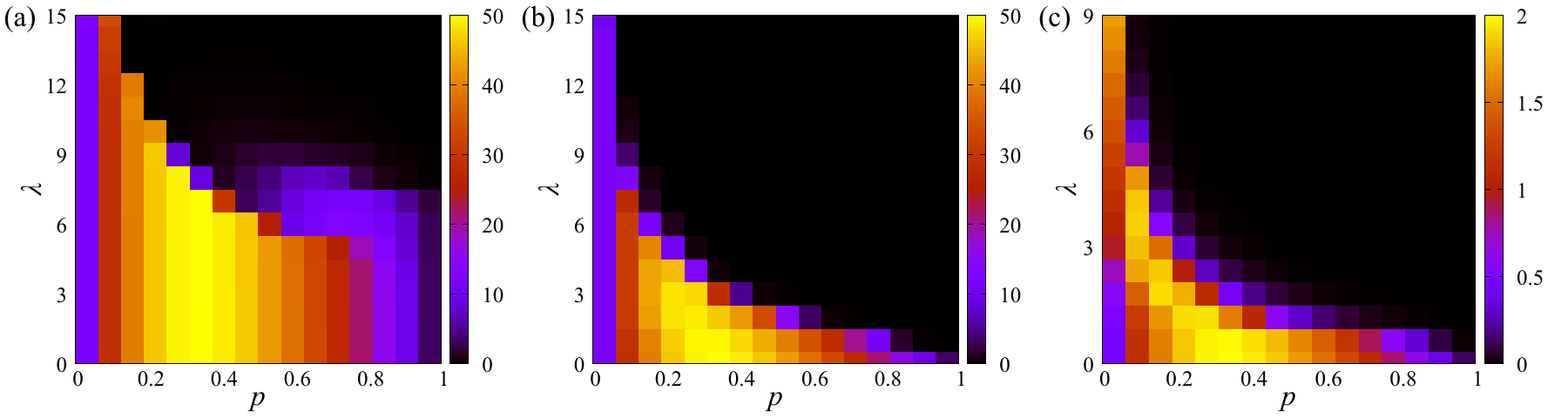}
\caption{
Frequency order parameter $F$ as a function of the degree of decentralization $p$ and the coupling strength $\lambda$.
For each value of $p$, simulations are performed by sweeping $\lambda$.
All results are averaged over 100 independent realizations.
(a) Forward process with $\alpha = 0.5$ and $\beta = 0.1$.
(b) Backward process with $\alpha = 0.5$ and $\beta = 0.1$.
(c) Forward process with $\alpha = 0.5$ and $\beta = 0.5$.
}
\label{Fig:Basic}
\end{figure*}

\vspace{0.1cm}
{\em Results}\textemdash
To characterize how coupling and decentralization jointly control frequency synchronization, we sweep the coupling strength $\lambda$ for each value of $p$ and record the steady-state value of $F$. Repeating this procedure over many random realizations of generator placements yields a synchronization diagram in the $(p,\lambda)$ plane, shown in Fig.~\ref{Fig:Basic}. This figure shows the frequency order parameter $F$ as a function of the coupling strength $\lambda$ for different degrees of decentralization $p$ with $\alpha = 0.5$ and $\beta = 0.1$.
Fig.~\ref{Fig:Basic}(a) corresponds to the forward process, in which $\lambda$ is increased from $\lambda = 0$ with an increment of $\Delta\lambda = 1$.
Fig.~\ref{Fig:Basic}(b) shows the backward process, where $\lambda$ is decreased with $\Delta\lambda = -1$, starting from $\lambda = 15$ (or from $\lambda = 60$ for $p = 1/32$ to $7/32$).

The key observation appears in the forward process: as the degree of decentralization $p$ increases, the coupling at which global frequency synchronization is reached ($F\to 0$) initially decreases, consistent with the expectation that distributing generation can facilitate synchronization. However, beyond intermediate $p$, this trend reverses: the required coupling increases again and a two-step approach to $F=0$ emerges for $p\simeq 0.6$--$0.8$. This non-monotonic behavior complements earlier studies reporting a monotonic reduction of the synchronization threshold under decentralization~\cite{2012RohdenSelf-Organized, Smith2022}, and highlights that the effect of decentralization can depend sensitively on how dynamical parameters such as inertia and damping co-vary with generation and consumption. Overall, decentralization can give rise to richer collective frequency-synchronization scenarios than a single-threshold picture suggests.

In particular, this phenomenon is not observed in the backward process,which adiabatically follows the synchronized branch (Fig.~\ref{Fig:Basic}(b)),  nor in the forward process with a relatively large regime $\beta$ (Fig.~\ref{Fig:Basic}(c)), where stronger dissipation reduces hysteresis and protocol dependence.

\begin{figure}[t]
\centering
\includegraphics[width=0.5\textwidth]{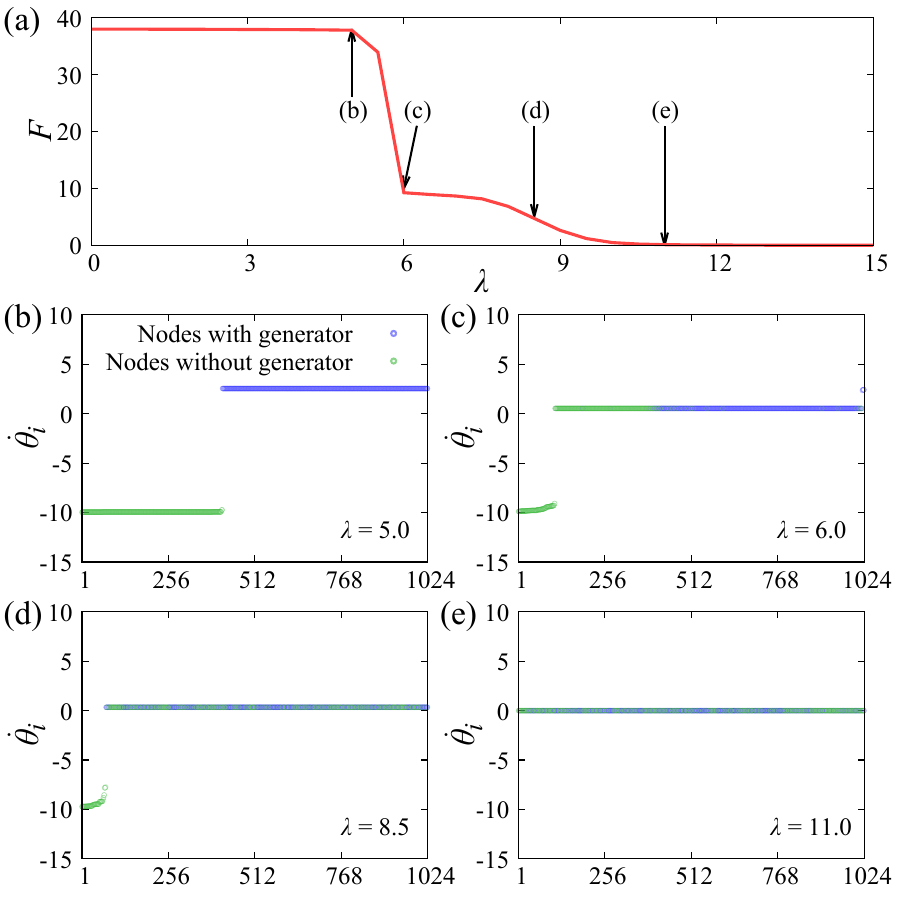}
\caption{
Forward process with $\alpha = 0.5$, $\beta = 0.1$, $p = 19/32$, and $\Delta\lambda = 0.5$.
(a) Frequency order parameter $F$ as a function of $\lambda$. Results are averaged over 1,000 independent realizations.
(b--e) Time-averaged frequencies of all 1,024 nodes in the stationary state for a single realization at each value of $\lambda$.
The nodes are ordered from lower to higher time-averaged frequency.
(b) $\lambda = 5.0$, (c) $\lambda = 6.0$, (d) $\lambda = 8.5$, (e) $\lambda = 11.0$.
}
\label{Fig:DoublePhaseTransition}
\end{figure}

\begin{figure*}[t]
\centering
\includegraphics[width=\textwidth]{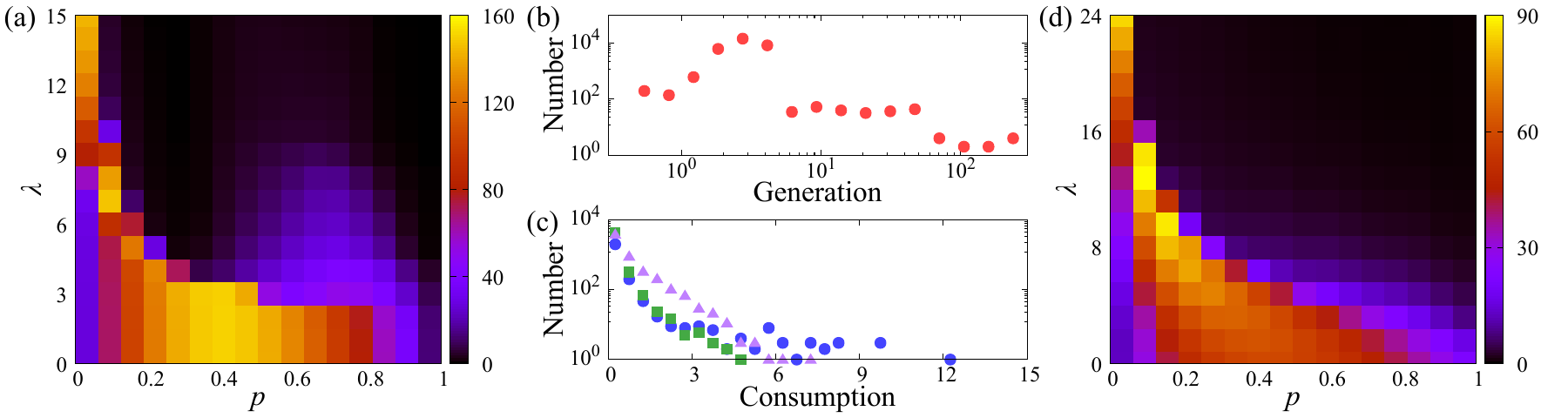}
\caption{
(a) Frequency order parameter $F$ as a function of the degree of decentralization $p$ and the coupling strength $\lambda$ in the forward process with $\alpha_c = 0.05$, $\beta_c = 0.05$, $\alpha_g = 0.5$, and $\beta_g = 0.1$.
(b) Distribution of rated power outputs of $30,757$ photovoltaic panels in the United Kingdom.
(c) Distribution of power consumption of $5,567$ households in London.
Consumption data are sampled at three randomly chosen time points between November 2011 and February 2014.
(d) Frequency order parameter $F$ as a function of the degree of decentralization $p$ and the coupling strength $\lambda$ in the forward process with $\alpha = 0.5$ and $\beta = 0.1$, using the empirical power-generation and power-consumption distributions shown in (b) and (c).
(a, d) For each value of $p$, simulations are performed by sweeping $\lambda$, and all results are averaged over 100 independent realizations.
}
\label{Fig:Real}
\end{figure*}

\vspace{0.1cm}
{\em Double phase transition}\textemdash 
Next, we examine in detail the double phase transition that emerges in the regime where the critical point increases with increasing decentralization $p$.
Fig.~\ref{Fig:DoublePhaseTransition}(a) shows a representative example of the double phase transition observed with $\alpha = 0.5$, $\beta = 0.1$, and $p = 19/32$.
As $\lambda$ increases, an abrupt transition occurs around $\lambda \simeq 6$, followed by a continuous transition near $\lambda \simeq 10$.

To elucidate how frequency synchronization develops as a function of $\lambda$, we examined the distribution of time-averaged frequencies of all nodes in the stationary state.
Before the first critical point, a clear bimodal distribution emerges, with nodes equipped with generators and nodes without generators forming distinct positive and negative frequency groups (Fig.~\ref{Fig:DoublePhaseTransition}(b)).
At the first critical point, a large fraction of nodes without generators abruptly synchronizes with the nodes equipped with generators, resulting in a discontinuous phase transition (Fig.~\ref{Fig:DoublePhaseTransition}(c)).
As $\lambda$ increases further, the remaining nodes without generators, which have not yet synchronized, gradually synchronize with the generator-equipped nodes (Fig.~\ref{Fig:DoublePhaseTransition}(d)).
After the second critical point, full frequency synchronization is achieved across all nodes (Fig.~\ref{Fig:DoublePhaseTransition}(e)).

\vspace{0.1cm}
{\em Robustness to realistic heterogeneity}\textemdash 
Finally, we examine whether our results remain robust when the model is modified to incorporate more realistic features.
First, consumer typically possess much smaller inertia than generator.
To account for this asymmetry, we assign inertia and damping coefficients to each node according to
\begin{equation}
\begin{aligned}
I_i &= \alpha_c |P_{i,c}| + \alpha_g |P_{i,g}|, \\
D_i &= \beta_c |P_{i,c}| + \beta_g |P_{i,g}|.
\end{aligned}
\end{equation}
Using the parameter values $\alpha_c = 0.05$, $\beta_c = 0.05$, $\alpha_g = 0.5$, and $\beta_g = 0.1$, we find that the double phase transition persists, as shown in Fig.~\ref{Fig:Real}(a).

Next, while keeping the inertia and damping coefficients defined as in the original model, we extract the distributions of power generation and consumption from empirical data.
Specifically, we use the distribution of rated power outputs from $30,757$ photovoltaic panels in the United Kingdom (Fig.~\ref{Fig:Real}(b))~\cite{2025EmpiricalGeneration}, normalized such that its mean equals 1, as the generation-output distribution in the model.
For power consumption, we employ the distribution of power usage from $5,567$ households in London, sampled at random times between November 2011 and February 2014 (Fig.~\ref{Fig:Real}(c))~\cite{2025EmpiricalConsumption}, normalized to have a mean value of $-1$.
For numerical stability, we imposed lower bounds on the inertia and damping constants: $I_i$ was set to 0.01 whenever its value from Eq.~\eqref{Eq:InertiaDamping} fell below 0.01, and $D_i$ was set to 0.002 whenever its value fell below 0.002.

In this case, the double phase transition is no longer sharply pronounced.
Nevertheless, we observe a broad regime in which the critical threshold for frequency synchronization remains nearly constant as the degree of decentralization $p$ increases.

\vspace{0.1cm}
{\em Conclusion}\textemdash In summary, we studied how decentralization affects frequency synchronization in power grids using a second-order Kuramoto model.
By introducing inertia and damping coefficients that scale with power generation and consumption, we relaxed the common assumption of uniform dynamical parameters.

Under this more realistic modeling framework, we show that decentralization does not always facilitate frequency synchronization. Although increasing decentralization initially lowers the critical coupling strength, further decentralization can increase it and induce a double phase transition.
We further demonstrated that the complex behaviors of critical threshold are robust when consumers are assigned smaller inertia and damping than generators.
Even when empirical distributions of photovoltaic generation and household electricity consumption are incorporated, we observe a region in which the critical threshold remains nearly constant as decentralization increases.
Our results highlight that decentralization alone does not guarantee improved frequency synchronization and emphasize the importance of realistic inertia modeling in the analysis of future power grids.

We acknowledge support from Spanish Ministerio de Ciencia e Innovacion (PID2021 128005NB-C21), Generalitat de Catalunya (2021SGR-00633) and ICREA Academia.
\bibliography{BibliographyMain}
\end{document}